\documentclass[12pt,preprint]{aastex}

\shortauthors {Kurtz et al}
\shorttitle {$\mu$--PhotoZ}

\begin {document}
\title{$\mu$--PhotoZ: Photometric Redshifts by Inverting the Tolman 
       Surface Brightness Test}

\author{Michael J. Kurtz, Margaret J. Geller, Daniel G. Fabricant, 
William F. Wyatt}
\affil{Harvard-Smithsonian Center for Astrophysics, Cambridge, MA 02138}
\and
\author{Ian P. Dell'Antonio}
\affil{Brown University, Department of Physics and Astronomy}
\affil{email: kurtz@cfa.harvard.edu}
\begin {abstract}

Surface brightness is a fundamental observational parameter of
galaxies.   We show, for the first time in detail, how it can be used
to obtain photometric redshifts for galaxies, the $\mu$-PhotoZ method.

We demonstrate that the Tolman surface brightness relation, $\mu
\propto (1+z)^{-4}$, is a powerful tool for determining galaxy
redshifts from photometric data.

We develop a model using $\mu$ and a color percentile (ranking)
measure to demonstrate the $\mu$-PhotoZ method.  We apply our method
to a set of galaxies from the SHELS survey, and demonstrate that the
photometric redshift accuracy achieved using the surface brightness
method alone is comparable with the best color-based methods.

We show that the $\mu$-PhotoZ method is very effective in determining
the redshift for red galaxies using only two photometric bands.  We
discuss the properties of the small, skewed, non-gaussian component of
the error distribution.

We calibrate $\mu_r, (r-i)$ from the SDSS to redshift, and tabulate
the result, providing a simple, but accurate look up table to estimate
the redshift of distant red galaxies.

\end {abstract}

\section {\label{intro} Introduction}

\citet{1930PNAS...16..511T} and \citet{1935ApJ....82..302H} first
showed that in an expanding universe the surface brightness of a
galaxy is a strong function of the redshift, SB $ \propto
(1+z)^{-4}$, or $\mu = \mu_0 + 10
log(1+z)$. \citet{1935ApJ....82..302H} suggested that this effect
could be used as a distance indicator and as a test of various
cosmological scenarios (the Tolman test).

\citet{1957AJ.....62....6B} was the first to show that the shifting of
spectral features as a function of redshift (in particular the
4000\AA\ break) causes color changes which can be calibrated to
estimate a redshift.  Interestingly the discussion with Hoyle
following \citet{1962IAUS...15..390B} demonstrates an understanding
that this effect might be combined with or be complementary to the
surface brightness effect for redshift estimation.  At the time of the
discussion, the steady state hypothesis was still quite viable, and
thus the exact nature of the surface brightness effect was not known.
More recently \citet{2001AJ....121.2271S} and
\citet{1996ApJ...456L..79P} measured the effect; various systematics,
including galaxy luminosity evolution, prevent these measures from
being definitive, but the true relation is very close to $(1+z)^{-4}$.

Typically, photometric redshifts are derived from color differences as
a function of total magnitude.  Some techniques fit model spectra
\citep[e.g.][] {1957AJ.....62....6B,
2001AJ....122.1163B,2000A&A...363..476B}, others directly calibrate
the dataset \citep[e.g.][]{2003MNRAS.339.1195F, 1999MNRAS.302..152K,
2006MNRAS.372..565F}, and still others combine these two methods into
a hybrid \citep[e.g.][P05]{2005MNRAS.359..237P}.  Several proponents
of the neural network method remark that surface brightness related
measures could be added to their ingest parameters
\citep[e.g.][]{2005AAS...207.2627M,2005AJ....130.2439S}.  By including
the Petrosian 50\%\ and 90\%\ flux radii \citet{2005PASP..117...79W}
found a 15\%\ improvement in the mean redshift error.

Here we demonstrate that surface brightness provides a highly
effective redshift estimator for the reddest galaxies in each surface
brightness interval, yielding very accurate photometric redshifts from
either $r$ and $i$, or $r$ and $z$ band observations.  We indicate how
other color combinations might be used. We present the $\mu_r,(r-i)$
calibration to redshift for red galaxies in the SDSS.  One powerful
advantage of this method is that only two bands are required for its
application.

We present the data in section \ref{data} and in section \ref{technique} we
introduce the $\mu$-PhotoZ technique, and discuss its physical basis
and limitations. We briefly indicate the extension of our technique to
other bandpasses in section \ref{colors}.  The technique is applied in
section \ref{direct} and we discuss the nature of the error distribution in
detail.  In section \ref{SDSS-calib} we apply our method to the SDSS.

\section {\label{data}Data}

We combine data from three large surveys to create a set of measures
suitable for demonstration of the {$\mu$}Photo-z technique; (i) a
magnitude limited galaxy catalog from the Deep Lens Survey
\citep[][DLS]{2006ApJ...643..128W}, (ii) redshifts and spectral types
from the Smithsonian Hectospec Lensing Survey
\citep[][SHELS]{2005ApJ...635L.125G}, and (iii) photometry and
spectroscopy from the fourth and fifth data releases of the Sloan
Digital Sky Survey \citep[][SDSS]{2006ApJS..162...38A}.

\subsection {DLS}

SHELS used the DLS deep R band photometry from the $2^\circ \times
2^\circ$ field F2 (9:20+30:00) to select galaxies for redshift
measurements with the Hectospec \citep{2005PASP..117.1411F}. We used
the difference between magnitudes within the 1.5$\arcsec$ and
5$\arcsec$ apertures and the FWHM to a gaussian fit as star/galaxy
classifiers.  It is possible to derive photometric redshifts based on
{\it colors} from the deep multicolor DLS data (Margoniner, et al, in
preparation).

The extended (scattering) halos around bright stars significantly
contaminate the brightness measures of faint galaxies.  We have
excluded all objects within a magnitude determined radius of every
bright star; about 5\%\ of the total area of the survey is thus
removed from further consideration.

\subsection {SDSS}

From the SDSS we use the Petrosian $r$ magnitudes \citep[for a
discussion of the advantages of Petrosian magnitudes for cosmological
investigations see][]{2002AJ....124.1810S}, the star/galaxy
classifications, and the colors derived from the fiber magnitudes.
From the Petrosian half light radii (in the $r$ band) we calculate the
central surface brightness $\mu_r$, the average surface brightness (in
$r$ mag per arcsec squared inside the half light radius); $$\mu_r
\equiv petroMag\_r + 2.5(0.798 + 2 log(petroRad50\_r))$$ All measures
are extinction corrected.

In section \ref{SDSS-calib} we calibrate the $\mu_r,(r-i)$ plane to
redshift for the SDSS, using $\mu_r,(r-i)$ and redshift.

\subsection {SHELS}

The SHELS will be a magnitude limited redshift survey when complete.
We use the data set as of 1 January 2006.  Our data set is 90\%\
complete to $R<19.7$ and is 50\%\ differentially complete at $R=20.3$,
using DLS total magnitudes.

From the full SHELS dataset we eliminated all objects in the vicinity
of bright stars (about 5\%); we discard any object where the DLS
position differs from the SDSS position by more than 1$\arcsec$; we
use only galaxies with spectra with secure redshifts.  The
final sample contains 8529 redshifts of the estimated final 12,000.

\subsection {The Combination}

The key dataset required for the $\mu$-PhotoZ technique is a
photometrically complete sample of galaxies with accurate central
surface brightnesses and at least one color.  For this paper we
primarily use $\mu_r$ and the $r-i$ color from the SDSS-DR4.  In
addition we derive a set of pseudo-magnitudes: $\mu_g$ =
$\mu_r+(g-r)$; $\mu_i$= $\mu_r-(r-i)$; $\mu_z$ = $\mu_r-(r-z)$, these
are estimates of the surface brightness inside the half-light radius
defined by the $r$ band.

\section  {\label{technique}Technique}

Surface brightness, which, in an expanding universe changes as
$(1+z)^{-4}$ \citep{1930PNAS...16..511T}, is a much more sensitive
indicator of redshift than apparent magnitude. Unlike apparent
magnitude, surface brightness has no dependence on the details of the
cosmology \citep[e.g.][]{1975gaun.book..761S}.

\begin{figure}
     \plotone{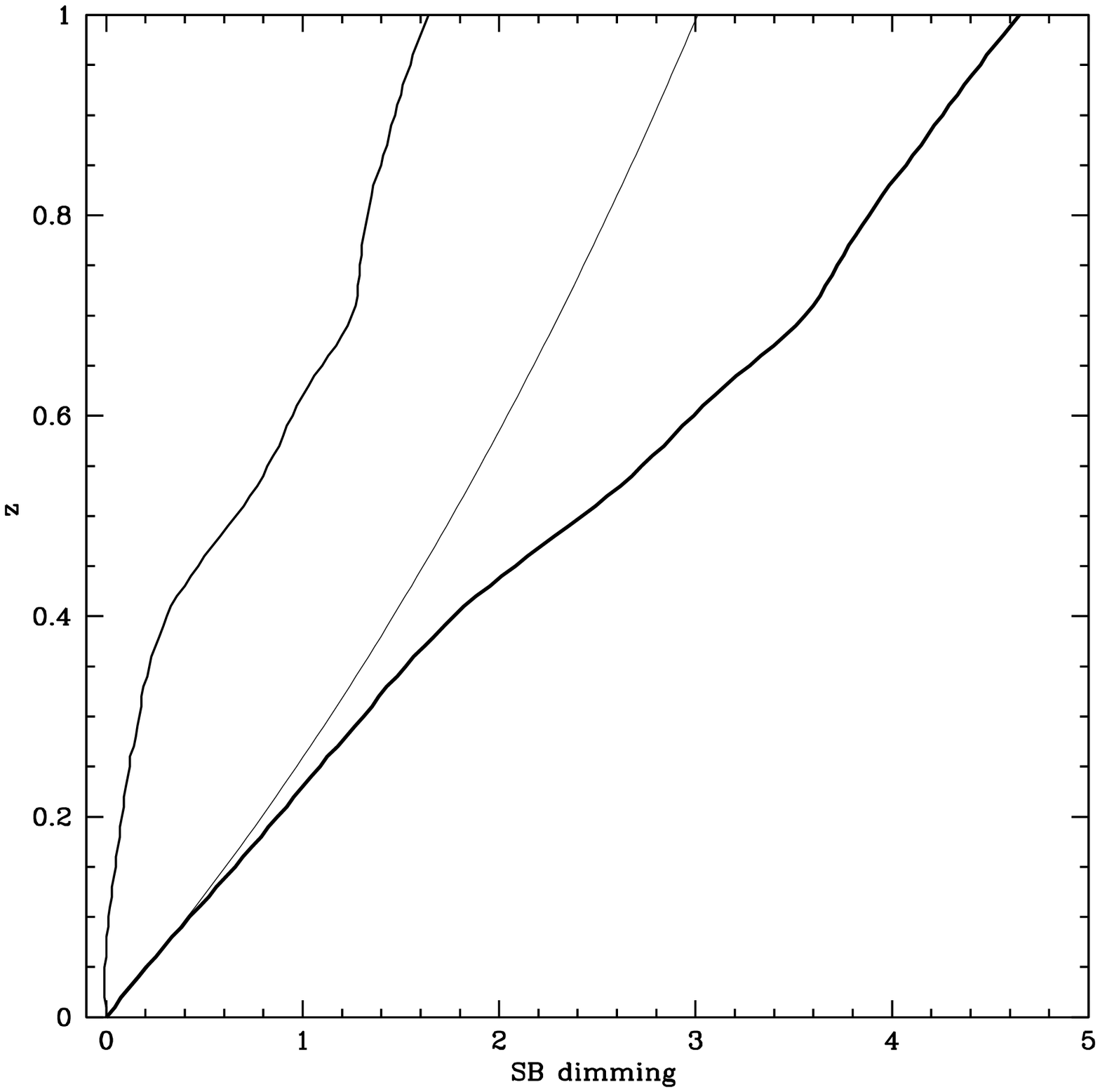}
     \caption{\label{SBdimming}
     Combining the K correction with surface brightness dimming, in
     magnitudes, as a function of redshift.  The
     \citet{2001annis} $r$ band BCG K correction is on the left, 
     the thin line in the
     center is $10 log(1+z)$, and the thick line on the right is the
     sum of the two effects.}

     \end{figure}

The change in the observed surface brightness of a galaxy is the
product of the change in brightness due to the use of a static
(non-redshifted) bandpass \citep[the K
correction][]{1936ApJ....84..517H} and the cosmological dimming.
Figure \ref{SBdimming} shows these effects in terms of the measured
dimming in magnitudes as a function of the redshift of the object.
The line on the left is the K correction for $r$ band measurements of
brightest cluster galaxies calculated by \citet{2001annis} using the
Pegase code \citep{2002A&A...386..446L}.  The thin line in the middle
is simply $10 log(1+z)$, and the thick line on the right is the product
of the two effects.

When the galaxies in the catalog are binned according to $\mu_r$, and
then sorted by color, the reddest 10-20\%\ of the galaxies in each
$\mu_r$ bin show a very strong correlation between $\mu_r$ and
redshift.

\begin{figure}
     \plotone{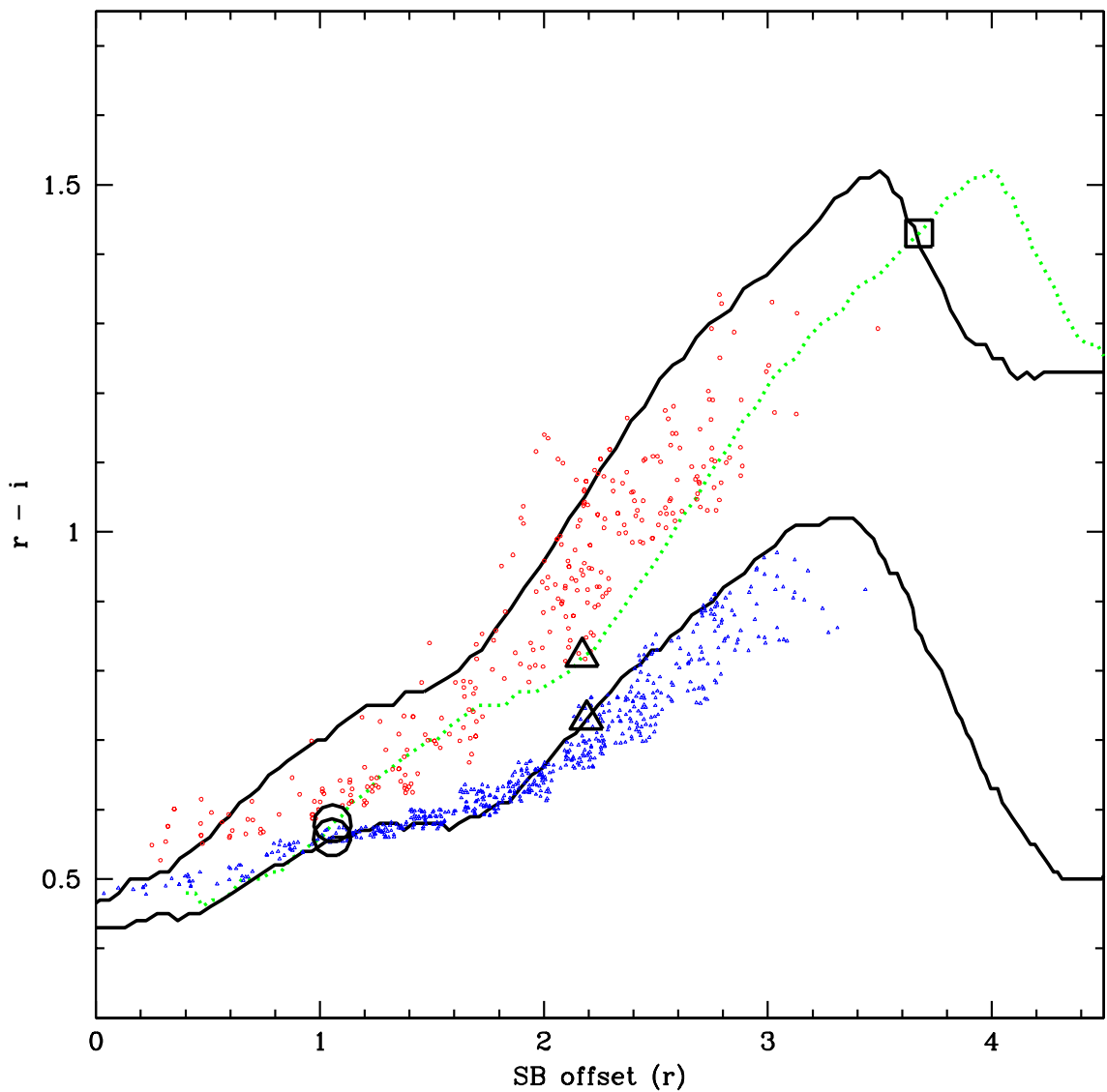}
     \caption{\label{types}
     Color vs. surface brightness dimming for different galaxy types.
     The x axis is the \citet{2001annis} $r$ band K correction plus
     $(1+z)^{-4}$ and the y axis is the difference between the 
     the K corrected $r$ band and $i$ band.   Redshift
     increases along the lines.  The left most lines represent the BCG
     K correction, with the right (green on-line) one shifted by 0.5
     mag with respect to the other.  The right most line is the Sa K
     correction.  The points represent the (red) reddest 10\%\ and
     (blue) the reddest 20--30\%\ of galaxies in the SHELS sample.
     The large geometric figures point to interesting regions
     described in the text.}

     \end{figure}

Figure \ref{types} explains this correlation.  The y axis is simply
the $r-i$ color.  The x axis represents the surface brightness dimming
expected from the sum of the appropriate K correction and the $10
log(1+z)$ effect.  The lines on the plot represent the $r-i$ color of
galaxies computed by subtracting the \citet{2001annis} K correction
model in the $i$ band from the model in the $r$ band.  On the left the
two lines represent the BCG galaxy model (black) and that model
shifted 0.5 mag (dashed green) to approximate the extent of the scatter.  On
the right, figure \ref{types} shows the model for an Sa galaxy
(black), we have shifted the zero point of the Sa relation by 0.46 mag
to account for the difference in $\mu_r$ between a $r^{1\over{4}}$ law
galaxy and an exponential disk with the same scale length and total
magnitude \citep{2002AJ....124.1810S}.  Redshift increases along the
lines.

The small red circles  show the observed $r-i$ color
and $\mu_r - const.$ for those galaxies in the reddest 10\%\ in each
$\mu_r$ bin which have a measured redshift, and the small blue
triangles show the 20-30\%\ reddest objects in each
bin.  As expected, the redder objects lie mainly between the two
BCG lines, and the bluer objects approximately follow the SA line,
validating our use of these models.

A successful technique which calibrates red galaxies to redshift must
be free of contamination by higher redshift blue objects at the same
color.  Figure \ref{types} shows that surface brightness eliminates
much of the possible confusion.  There are, however, regions in figure
\ref{types} where lines cross, or come close to each other, indicating
degeneracies.

There are three sets of large geometric figures on the plot: the left
(bottom) two (circles and triangles) show regions where there could be
confusion between BCG galaxies and Sa galaxies, indicating possible
problems for the technique; the right (top) square(s), which overlap
perfectly, show where BCG galaxies no longer become redder in $r-i$
with increasing redshift, indicating a limit to the $\mu$-PhotoZ
technique.

The circles on the bottom left show where the BCG galaxy
is at z=0.15 and the SA galaxy is at z=0.24.  Clearly if a calibration
of red objects to redshift as a function of $\mu_r$ succeeds at this
redshift it is not because it separates early type galaxies from
SA galaxies.

The pair of triangles are at z=0.4 for the BCG and z=0.5 for the
SA. They are close enough to present a potential problem.  We note
that P05 found increased scatter in their calibration of photometric
redshifts for LRG galaxies at z=0.4, but attributed it to other
causes.  Notice that it is the (approximately) 0.45 magnitude
difference between the BCG and Sa central surface brightnesses at a
fixed magnitude which separates the types in the color-magnitude
diagram in figure \ref{types}; previous methods, such as P05, use
total magnitudes and additional colors.  Using surface brightness
breaks this degeneracy for galaxies with different surface brightness
profiles, without requiring additional colors be observed.

The top square(s), at z=0.65 for the unshifted BCG locus and z=0.765
for the shifted locus clearly shows a region where the $10 log(1+z)$
dimming is insufficient to remove degeneracies.

To summarize the technique: using a complete photometric catalog
(e.g. SDSS) we (1) bin galaxies according to a measure of central surface
brightness, such as $\mu_r$, then (2) in each bin, we sort the galaxies by a
measure of color, such as $r-i$.  Next (3) we assign a percentile rank to
each galaxy based on its color, within each surface brightness
bin. 

As an example, in the F2 field there are 2506 galaxies with $\mu_r$
between 22.6 and 22.8; a galaxy at $\mu_r= 22.79$ and with $r-i =
0.965$ is redder than 96.2\%\ of the galaxies in its bin, and
receives a color percentile of 96.2.  Likewise a galaxy with $\mu_r =
22.38$ and with $r-i = 0.769$ is redder than 96.2\%\ of the 1551
galaxies in the 22.2-22.4 $\mu_r$ magnitude bin.  Note that both
galaxies have a color rank score of 96.2 but have substantially
different actual $r-i$ colors.

These color rank percentile (CRP) scores are strongly correlated with
galaxy type; in the next section we demonstrate their usefulness in
permitting a calibration of central surface brightness to redshift for
the redder objects.

\subsection {\label{results}Results}

\begin{figure}
     \plotone{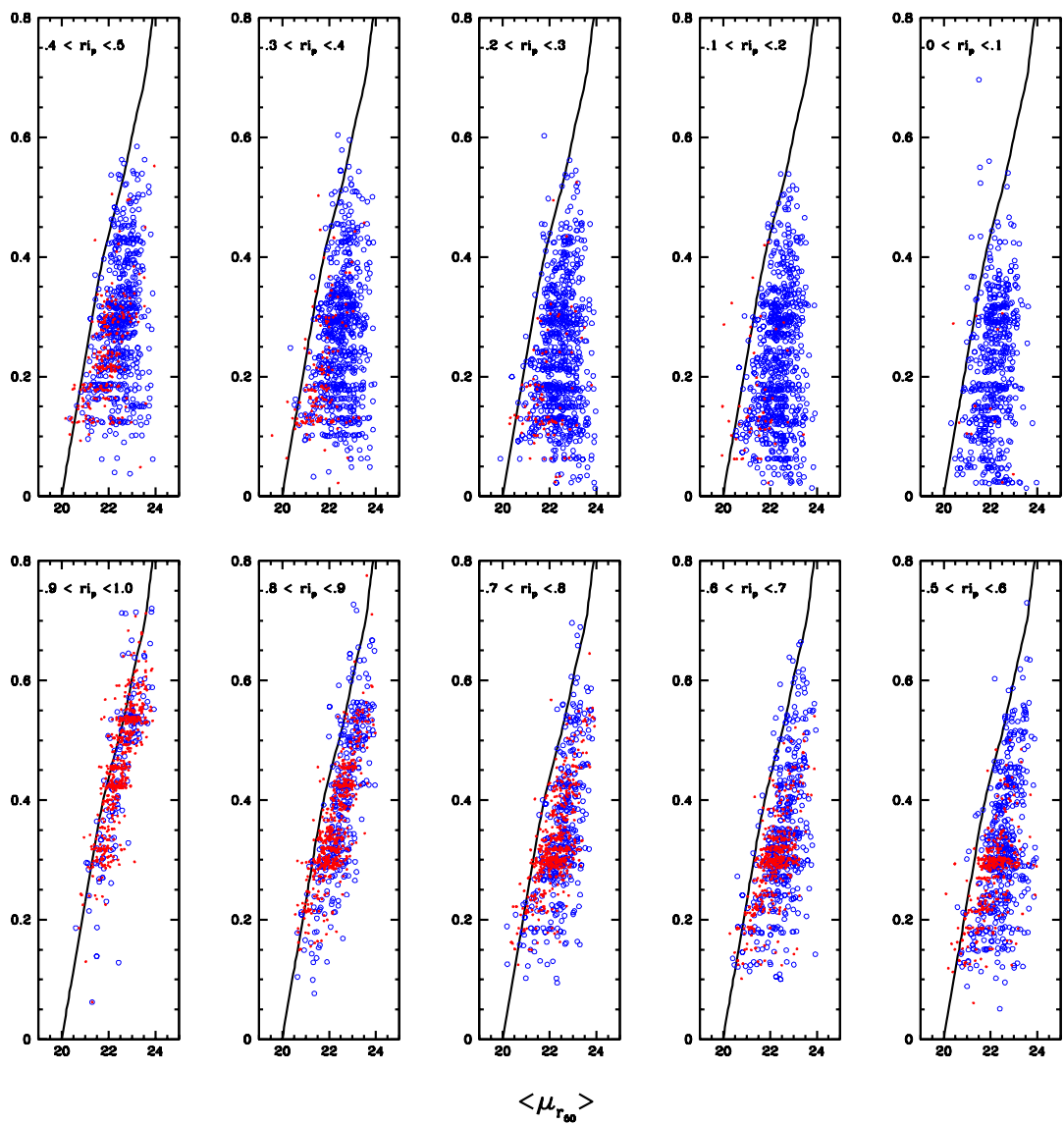}
     \caption{\label{10panel.p50r}
     The redshift-surface brightness diagram for ten deciles in $r$
     surface brightness sorted $(r-i)$ color (Color Rank Percentile, see text).
     The (online blue)
     circles are objects with clear O[II] emission, the (red)
     dots are absorption line objects.}

     \end{figure}

Figure \ref{10panel.p50r} shows the result of plotting $\mu_r$
vs. redshift for galaxies within ten decile bins in CRP
scores.  We compute ranks using the $r-i$ color.  The red dots are for
spectra where the measured O[II] 3727\AA\ equivalent width is
$>-2$\AA\ and the blue circles have clearly measurable emission,
($EQW_{3727}<-2$\AA\ ) .  The solid lines are the \citet{2001annis}
$r$ BCG K correction times $10 log (1+z)$ shifted to match $\mu_r =
20.0$ mag per square arcsec at redshift zero.

There are several things to notice in figure \ref{10panel.p50r}.
First the galaxies in the reddest deciles in CRP follow the prediction
of the \citet{2001annis} BCG K correction times $10 log (1+z)$ very
closely. The next decile is slightly offset from the reddest.
Clearly the calibration of $\mu_r$ to redshift for galaxies in these
CRP deciles is straightforward.

Next we notice that the bluest objects show essentially no correlation
between $\mu_r$ and redshift, the available $\mu_r - z$ space in the
bluest decile in CRP is nearly uniformly filled. Clearly $\mu_r$
cannot be calibrated to redshift for the bluest objects, which have
their surface brightnesses altered by star formation.

\begin{figure}
     \plotone{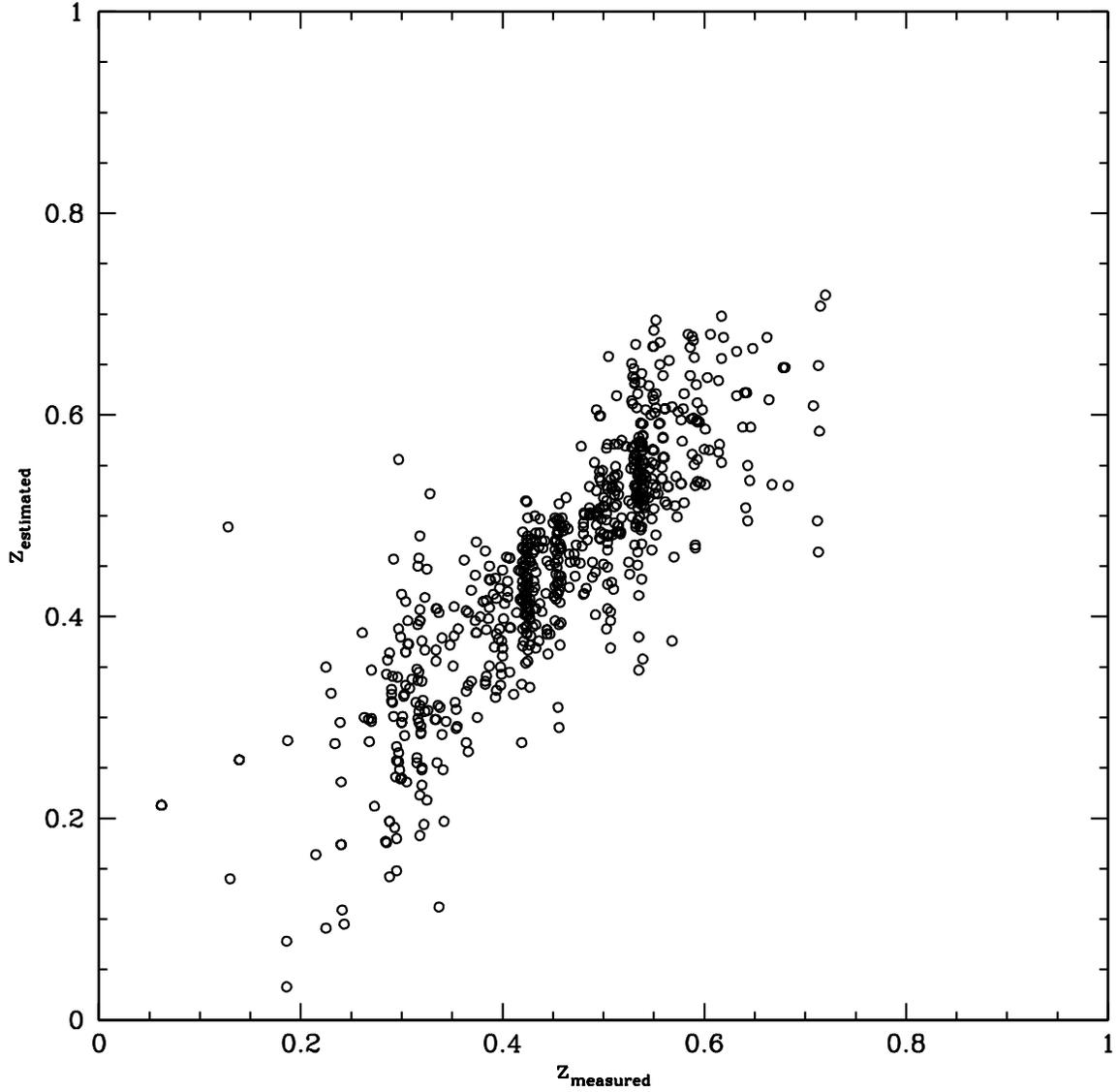}
     \caption{\label{resid.fit.9}
     A comparison of the predectided and measured redshifts obtained by
     applying the model to the reddest 10\%\ of galaxies. }

     \end{figure}

Figure \ref{resid.fit.9} shows the result of fitting $\mu_r$ to
redshift by finding the redshift, $z$, where for the $i^{th}$ galaxy,
$$\mu_r(i) = K_{i}(z) + 10 log(1+z) + 19.45 + 3.3(1-CRP(i))$$ (the
model).  The factor containing CRP accounts for the change in $\mu_r$
with color, variously quantified as the color-magnitude relation for
elliptical galaxies
\citep[e.g.][]{1959PASP...71..106B,1977ApJ...216..214V,
2006MNRAS.372..199C,2006AJ....131..736C,2006astro.ph.11873E} or as the
fundamental plane \citep[e.g.][]{1987ApJ...313...59D,
2003AJ....125.1866B}.

The standard deviation for the entire sample of the reddest 10\%\ in
figure \ref{resid.fit.9} is $\sigma({\Delta z/{1+z})}=0.046$ and is
$\sigma=0.056$ for the reddest 10--20\%.  This result is comparable
with the best current photometric redshifts for the LRG galaxies. P05
obtain $\sigma=0.035$ using $g$ photometry as well as $r$ and
$i$ (see section \ref{direct}).

\section {\label{colors}Using other colors and magnitudes}

As figure \ref{types} shows, the $\mu$-PhotoZ technique becomes
degenerate when the 4000\AA\ break enters the bandpass of the reddest
band ($i$ in figure \ref{types}).  To acquire $\mu$-PhotoZs for higher
redshift objects observations in a redder passband ($z$, J, H, etc)
 would extend the technique to higher redshifts.  Neither the
SHELS spectroscopy nor the SDSS photometry go deep enough to test the
technique for redshifts greater than 0.65 using $z$ band photometry.

\begin{figure}
     \plotone{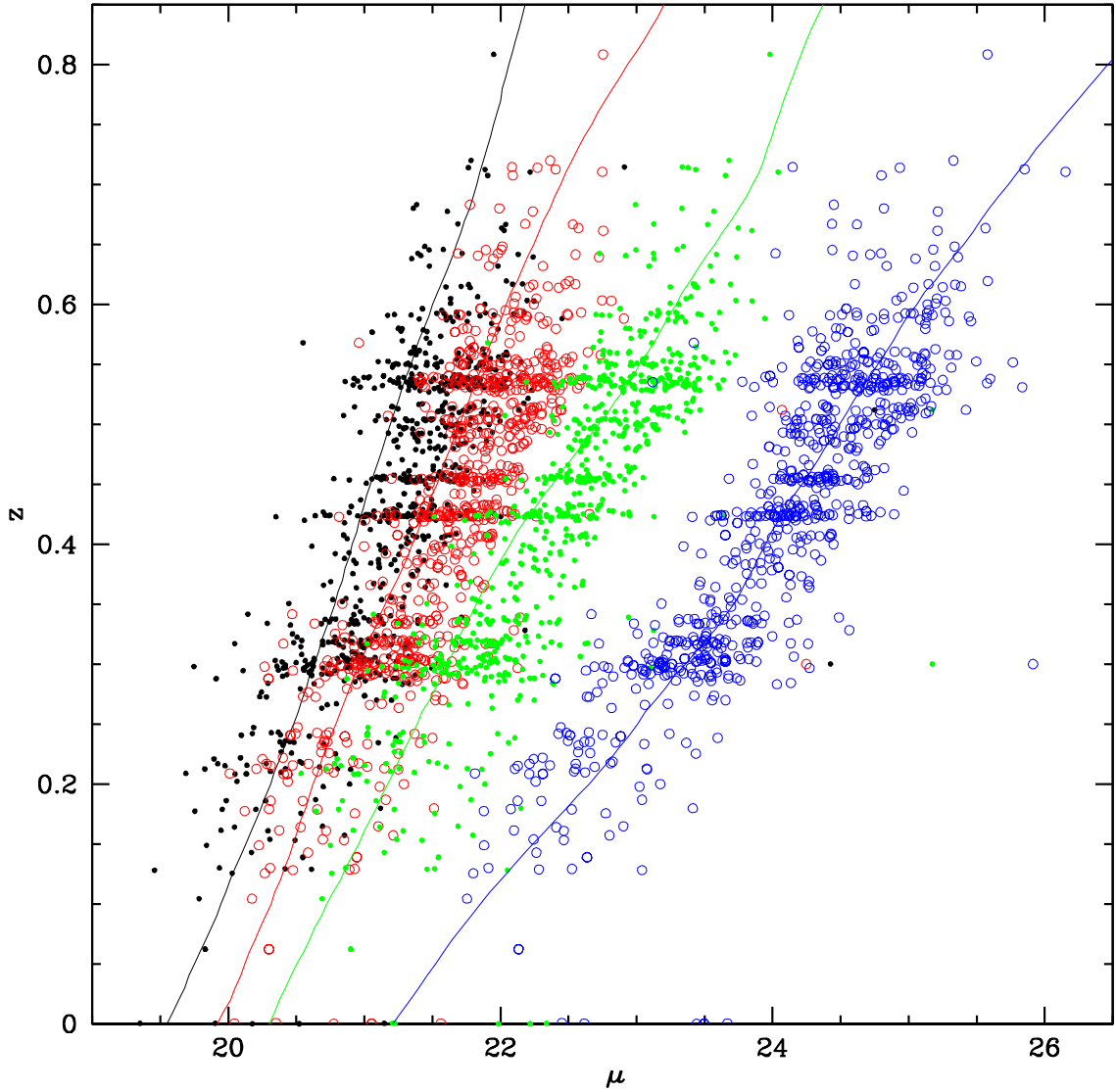}
     \caption{\label{4color.mu}
     The surface brightness redshift relation for different color
     bands, for the reddest 10\%\ in $(r-i)$ selected at fixed
     $\mu_r$.  The leftmost (black online) dots are the $z$ band measure,
     the (red) circles are $i$, the (green) dots are $r$,
     and on the rightmost the (blue) circles are $g$.  The lines are the
     \citet{2001annis} BCG K correction for the appropriate color.}

     \end{figure}

Obviously other bandpasses may also be used.  There are three different
choices to make: (1) the bandpass to use to sort the colors, (2) the
color to use, and (3) the bandpass to fit.  Here we have used $\mu_r,
(r-i)$, and $\mu_r$, but other choices are informative.  Figure
\ref{4color.mu} shows galaxies in the the reddest 10\%\, as determined
using $\mu_r$ and $(r-i)$ versus $\mu_z, \mu_i, \mu_r, \mu_g$, along
with the sum of $10 log(1+z)$ and the appropriate \citet{2001annis}
BCG K correction for each color.  Redshift estimates using $z$ and $g$
band data have larger errors than estimates using $r$ and $i$, in the
manner expected from the larger published errors in the SDSS
magnitudes for these bands.

\begin{figure}
     \plotone{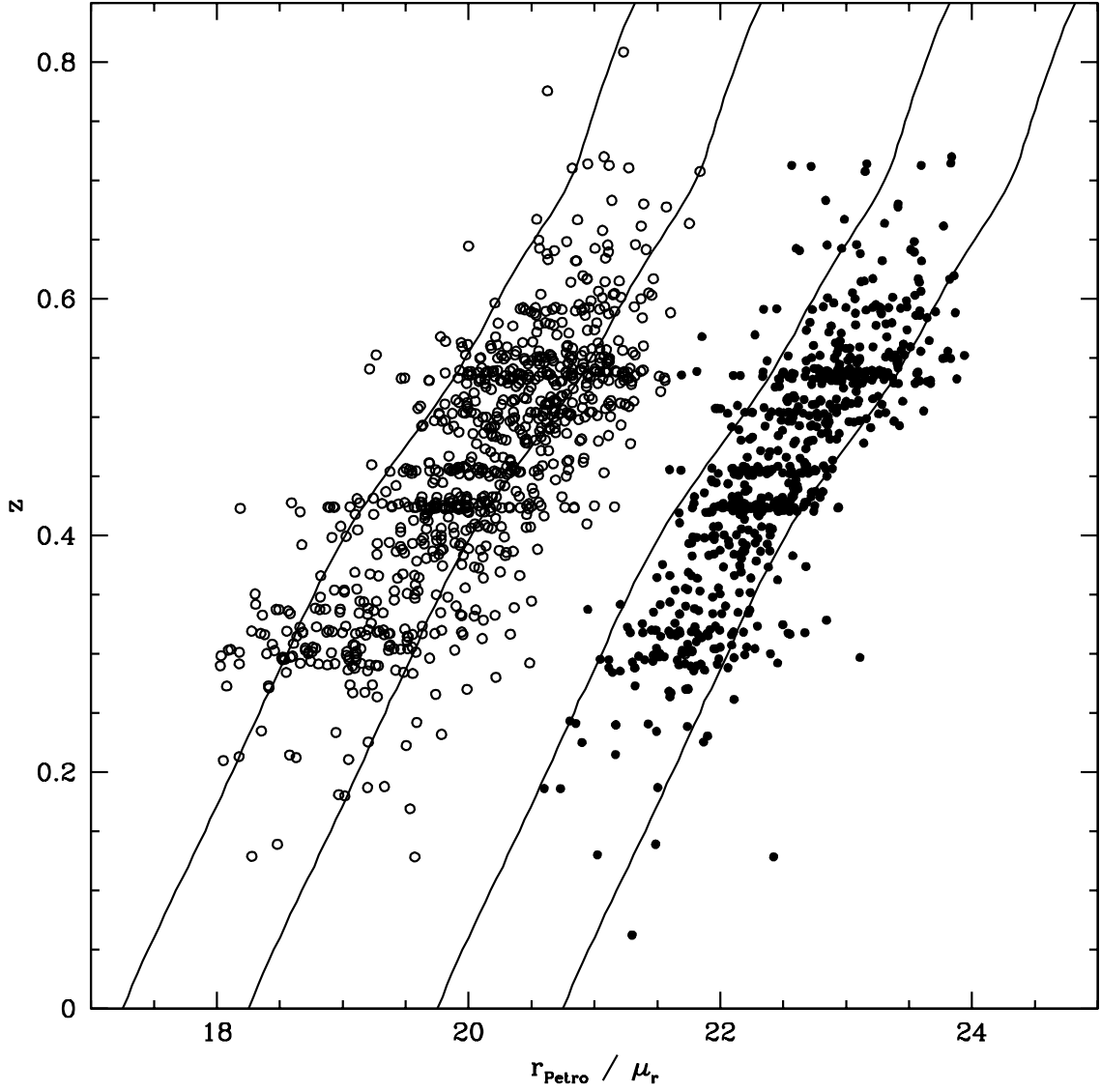}
     \caption{\label{petrovsmu}
     The difference between the reddest 10\%\ using total Petrosian
     magnitudes (left) and $\mu_r$ (right).  Note the decreased scatter
     for $\mu_r$.}

     \end{figure}

One might also ask whether the technique of using color ranks might
also work if one uses total magnitude rather than central surface
brightness.  Figure \ref{petrovsmu} shows (left side) galaxies in the
the reddest 10\%\ in $(r-i)$ measured at fixed Petrosian $r$ magnitude
versus Petrosian $r$, and (right side) galaxies in the reddest 10\%\
in $(r-i)$ measured at fixed $\mu_r$ versus $\mu_r$; each grouping has
two lines to guide the eye. These lines are the product of the K correction
and the surface brightness dimming. The separation is an arbitrary
one magnitude.

Clearly the scatter in the Petrosian magnitude measures in figure
\ref{petrovsmu} is larger than in the $\mu_r$ measures. Figure
\ref{types} explains this behavior.  Recall that we shifted the line
representing the K correction and surface brightness dimming by 0.46
magnitudes to account for a 0.46 difference in the mean central
surface brightness at fixed total magnitude for the two galaxy types
(BCG and Sa), and this procedure matched the data.  Shifting the Sa line
0.46 magnitudes brightward would place it much closer to the BCG line,
and would thus produce substantially more degeneracy in the color rank
approach.

\section {\label{direct}Calibrating the Relation Directly}

In section \ref{results} we demonstrated that we can
predict the redshift of a red galaxy accurately from a measure of its
central surface brightness and a single color ($r-i$).  We fitted
models based on estimates of the K-correction, the color-magnitude
relation for elliptical galaxies, and the $(1+z)^{-4}$ cosmological
dimming to measures of color and surface brightness.

We may not have used the best model parameters in making these fits,
and we can also suspect that our measure of central surface
brightness, $\mu_r$ could have measurement systematics which correlate
with redshift.  Although these models are useful in understanding the
physical basis for the technique, they are not actually required to
implement it.

Rather than attempt to optimize the models and measures, we choose to
calibrate the data directly.  To each galaxy we assign the median
redshift of its neighbors in ($\mu_r$, CRP) space, excluding the
galaxy itself, in jack-knife fashion.  

\begin{figure}
     \plotone{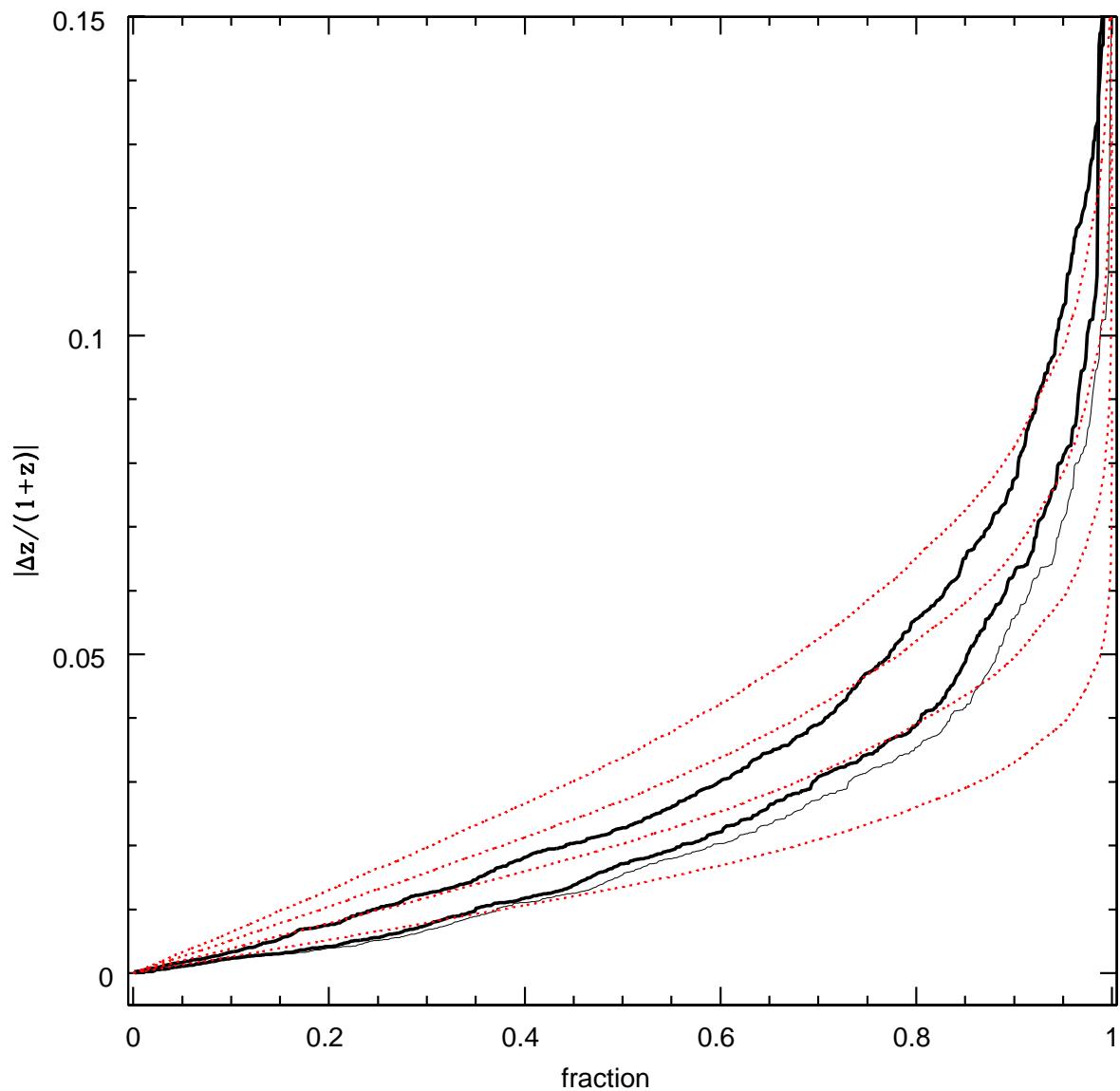}
     \caption{\label{p50r.fracvsgauss}
     The sorted absolute value of the errors of the technique in
     section \ref{direct}.  The (red online) dotted lines represent
     normal error distributions, with $\sigma$s of 0.05 (top), 0.04,
     0.03, and (bottom) 0.02.  The thick lines represent the actual
     data for the reddest 80-90\%\ (top) and the reddest 90-100\%\
     (bottom).  The thin solid line and the lower solid thick line differ by
     the removal of all objects
     with $(g-r)<1.4$ for the thin line.}

     \end{figure}

Figure \ref{p50r.fracvsgauss} shows the distribution of absolute
values of the residuals (in terms of $\Delta z/(1+z)$); the thick
black lines represent the residuals of the actual data (top represents
galaxies in the 80-90 color percentile range, bottom represents those
in the 90-100 color percentile range), sorted from smallest to
largest. The dotted lines show the distributions expected for normal
error distributions with $\sigma$s of (bottom to top) 0.02, 0.03,
0.04, and 0.05.  The tails of the error distribution are clearly not
gaussian, but the errors are small.  Eighty percent of all objects in
the reddest decile have errors smaller than would be expected for a
normal error distribution with a $\sigma$ of 0.03.

About 40\%\ of the high residual objects are low redshift galaxies
that are anomalously blue in $g-r$. These objects could easily be
removed in a three band survey.  For example, the thin solid line in
figure \ref{p50r.fracvsgauss} shows the error distribution for the
same sample as the bottom solid line, but with all objects with
$g-r<1.4$ removed.  The tails of the distribution are substantially
reduced.

\begin{figure}
     \plotone{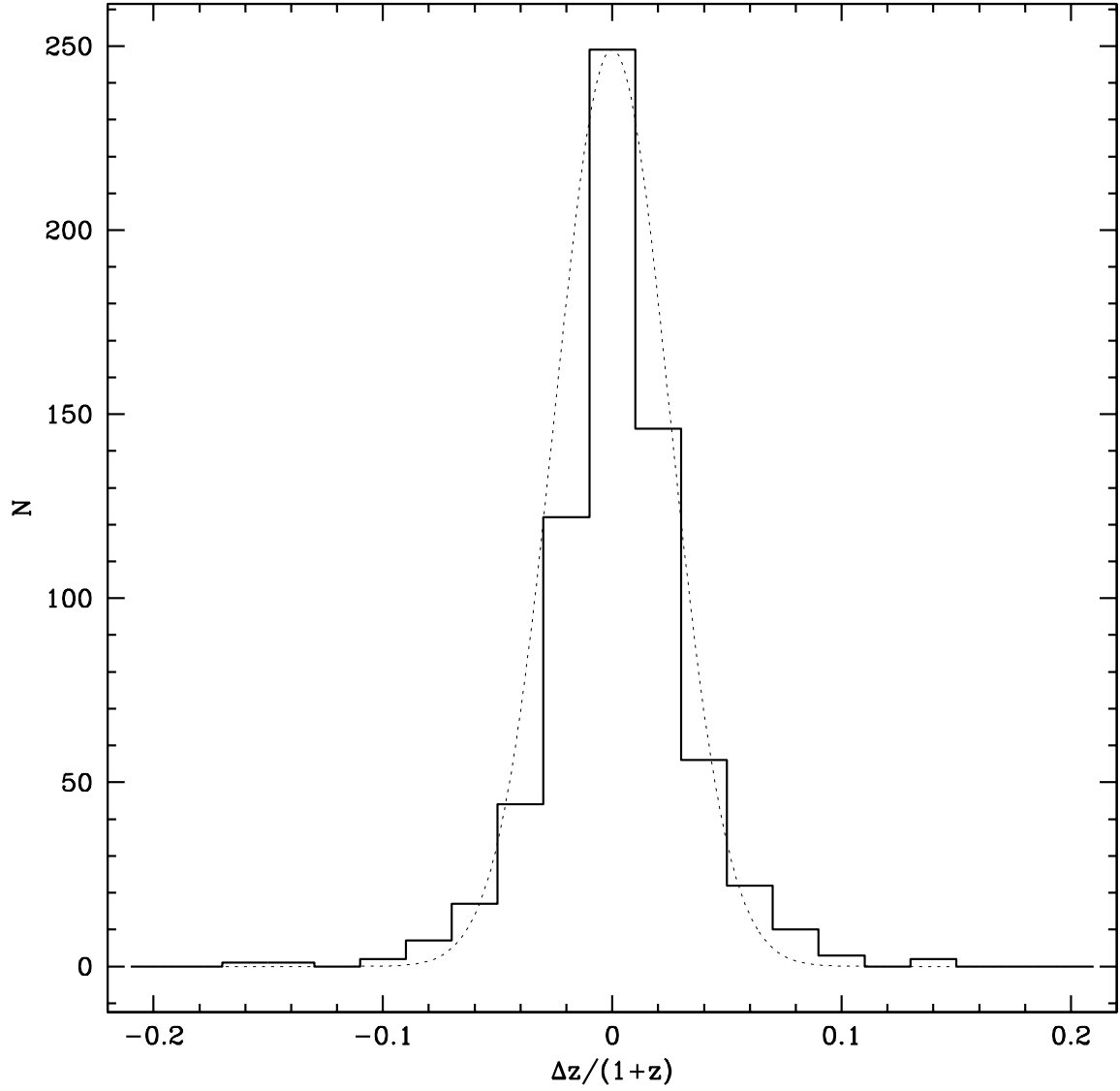}
     \caption{\label{p50r.Deltaz.histo}
     The error distribution ($\Delta z \over {1+z}$) for the 80-100\%\
     reddest objects, with the $(g-r)<1.4$ objects removed. The
     gaussian has $\sigma = 0.025$.}

     \end{figure}

Figure \ref{p50r.Deltaz.histo} shows the distribution of residuals for
the reddest 20\%\ (excluding $g-r<1.4$)
$(z_{measured}-z_{estimated})\over{1+z}$ with a $\sigma=$0.025
gaussian overplotted (1272 objects with redshifts).  These results show
that $\mu$-PhotoZ redshifts are as accurate as the best photometric
redshifts, (e.g. P05).  The 1$\sigma$ RMS for the data in figure
\ref{p50r.Deltaz.histo} is 0.030, P05 obtain 0.035.  The
$(\mu_r,CRP)-z$ calibration works well for all galaxies with
$(g-r)>1.4$. The 1$\sigma$ RMS for all these 2399 galaxies with
redshifts is 0.037.

\begin{figure}
     \plotone{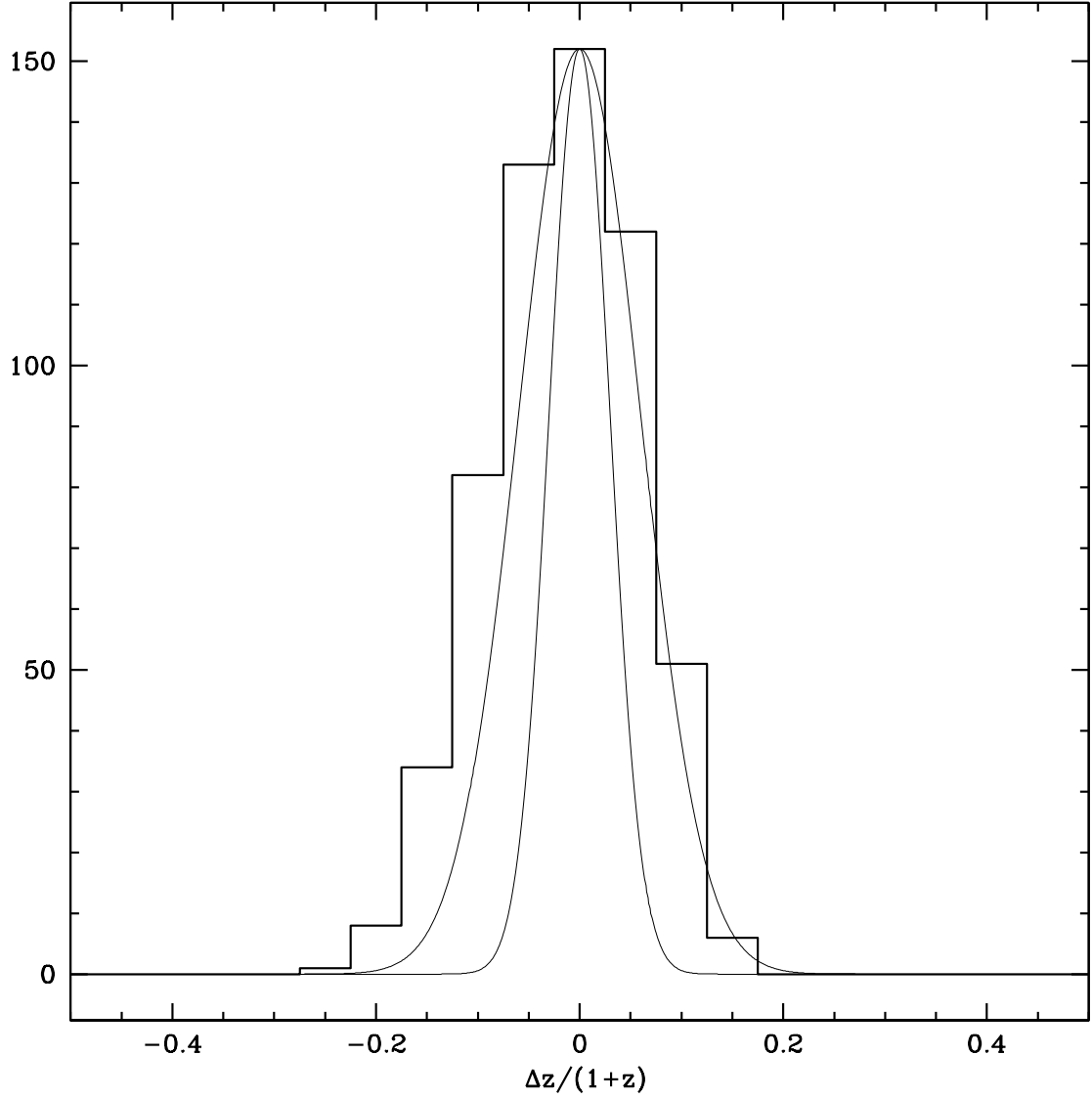}
     \caption{\label{bluegauss}
     The error distribution ($\Delta z \over {1+z}$) for the 80-100\%\
     reddest objects, with $(g-r)<1.4$.  Gaussians with $\sigma =
     0.030$ and $\sigma=0.06$ are superposed.  Note the change in
     scale from figure \ref{p50r.Deltaz.histo}}

     \end{figure}

Figure \ref{p50r.fracvsgauss} clearly indicates that the $\mu$-PhotoZ
method provides estimated redshifts from a single color, such as
$r-i$, albeit with increased errors and a non-gaussian error
component.  Figure \ref{bluegauss} shows the error distribution
$\Delta{z}\over{1+z}$ for the 330 galaxies from SHELS which are in the
reddest 20\%\ and which have $g-r<1.4$; two gaussians, with
$\sigma=0.03; 0.06$ are also plotted.

The error distribution in figure \ref{bluegauss} for
$\Delta{z}\over{1+z}$ can be reasonably fit by a gaussian with
$\sigma=0.06$ and a non-gaussian tail, where the actual redshifts are
always smaller than the estimated ones.  A description of the tail,
consistent with the current data, is that the objects are randomly
located in the galaxy distribution in the foreground of its apparent
(as estimated by $\mu$-PhotoZ) position, $z_{est}$.

For the entire sample of the 20\%\ reddest galaxies, without removing
any objects using a third band, the error distribution for $z_{est}$:
$$ Err_{z_{est}} = 0.79 G(.03) + 0.16 G(.06) + 0.05
R(N(z));(z<z_{est})$$ 

where $G(x)$ is a gaussian with $\sigma=x$; $R(x)$ is a random deviate
from the distribution $x$, and $N(z)$ is the redshift distribution of
the ``foreground'' galaxies.

\section {\label{SDSS-calib} Calibration the SDSS $\mu_r, (r-i)$ to 
redshift relation for red galaxies}

The color percentile measure at any fixed surface brightness
corresponds directly to an actual color.  To provide a calibration for
the SDSS we simply find the median redshift for galaxies in small
cells in $\mu_r, (r-i)$.

We use the SDSS-DR5 and find the 62,259 galaxies with
$20.35<\mu_r<23.25$ and $0.475<(r-i)<1.225$ and with $(g-r)>1.4$ which
have measured redshifts.  59,905 redshifts come from the SDSS,
essentially the LRG sample \citep{2001AJ....122.2267E}, and 2,345
redshifts come from SHELS, mainly for fainter galaxies.  There are 64
objects in common.  The redshifts agree, save for one case where SDSS
found a star and SHELS found a galaxy.

To estimate the error in the redshift estimator we take, for each
cell, the galaxies in the cell and assign to them the median redshift
for that cell.  We then rank the absolute value of the residuals for
the galaxies in that cell.  The median residual, for a gaussian
distribution would be $0.67\sigma$, the $68^{th}$ percentile would be
$1\sigma$ and the $95^{th}$ percentile would be $2\sigma$.  We only
estimate the $1\sigma$ point if there are at least 15 galaxies in a
bin, and the $2\sigma$ point if there are more than 30.

Table \ref{sdss-table} displays the results for bins with small and
well determined residuals in a convenient form.  The typeface
indicates the error. Use of this table allows quick and reasonably
accurate estimation of redshifts for red galaxies in the SDSS
photometric database reaching to a redshift of approximately 0.5.

\section {\label{discussion}Discussion}

Surface brightness is a fundamental observational parameter of
galaxies; it is directly related to redshift through the classical
Tolman effect.  We believe this investigation is the first to show in
detail that surface brightness may be used directly as a measure of
redshift, the $\mu$-PhotoZ method.

We demonstrate the value of surface brightness measures in determining
photometric redshifts using three different techniques.  First, we use
a model which emphasizes the physical and measurement aspects of the
problem.  Next, we demonstrate that by assigning a galaxy the median
redshift of its neighbors in surface brightness, color percentile
space we can achieve photometric redshift errors comparable with the
best current photometric redshift methods.  Finally by taking the
median redshift in small bins in surface brightness--color space we
develop a redshift estimator with errors comparable with the best
techniques.

Although we have concentrated on the use of $\mu_r$ and $(r-i)$ for
nearby red galaxies the techniques discussed here have much greater
generality; the $(1+z)^{-4}$ cosmological dimming effects all galaxies
at all redshifts.  Adding surface brightness directly into model
fitting and hybrid methods seems a very profitable avenue to explore.

Obviously the technique could be extended to greater redshift for red
galaxies by using the $z$ band.  Deep, large area lensing surveys,
including the DLS, often have deep $z$ photometry, and thus can obtain
photometric redshifts out to a redshift of about 1, using the methods
shown here.  These deeper surveys will also be able to eliminate some
of the non-gaussian error apparent in the SDSS photometry.  Much of
this error results from SDSS failure to separate close pairs of faint
objects (star + galaxy or galaxy + galaxy).  With much deeper imaging
and better seeing these effects should be less troublesome.

As the discussion in section \ref{direct} makes clear, most of the
benefit from applying the $\mu$-PhotoZ technique is achieved by using
just two photometric bands.  This powerful aspect of the technique has
obvious implications for the design of large area weak lensing
surveys.  It is exactly the high central surface brightness red
galaxies analyzed in section \ref{direct} which have the highest
redshift at fixed central surface brightness (figure
\ref{10panel.p50r}), and central surface brightness is the limiting
factor in determining the shape parameters for the sources.  Our
technique makes feasible substantially larger area weak lensing
surveys than would be possible were it necessary to observe through
several filters.

\section{Acknowledgments}

We thank Scott Kenyon for discussions, and we thank Tony Tyson, David
Wittman and Vera Margoniner for their collaboration in obtaining the
initial galaxy catalog.  We thank Ken Rines and Warren Brown for a
careful reading of the manuscript.  This research was supported in part
by the Smithsonian Institution.

{\it Facilities:} MMT(Hectospec), SDSS, KPNO(Mayall).


    \clearpage

\clearpage

\begin{deluxetable}{lccccccccccccccc}
\tabletypesize{\tiny}
\tablecaption{Redshift vs. $\mu_r$ and $(r-i)$ \label{sdss-table}}
\tablecolumns{16}
\tablewidth{0pt}

 \tablehead{\colhead{$\mu_r$}&
 \colhead{0.50}&
 \colhead{0.55}&
 \colhead{0.60}&
 \colhead{0.65}&
 \colhead{0.70}&
 \colhead{0.75}&
 \colhead{0.80}&
 \colhead{0.85}&
 \colhead{0.90}&
 \colhead{0.95}&
 \colhead{1.00}&
 \colhead{1.05}&
 \colhead{1.10}&
 \colhead{1.15}&
 \colhead{1.20}}
\startdata
 20.40&
 &           
 &           
 &           
 &           
 &           
 &           
 &           
 &           
 &           
 &           
 &           
 &           
 &           
 &           
 {\tt      }
 \\
 20.50&
 &           
 &           
 &           
 &           
 &           
 &           
 &           
 &           
 &           
 &           
 &           
 &           
 {\tt      }&
 {\tt      }&
           
 \\
 20.60&
 {\tt 0.228}&
 {\tt 0.258}&
 &           
 &           
 &           
 &           
 &           
 &           
 &           
 &           
 &           
 &           
 &           
 &           
 {\tt      }
 \\
 20.70&
 {\bf 0.246}&
 {\rm 0.255}&
 &           
 &           
 &           
 &           
 &           
 &           
 &           
 &           
 &           
 &           
 {\tt      }&
 {\tt      }&
 {\tt      }
 \\
 20.80&
 {\bf 0.235}&
 {\bf 0.265}&
 &           
 &           
 &           
 &           
 &           
 &           
 &           
 &           
 &           
 &           
 {\tt      }&
 &           
 {\tt      }
 \\
 20.90&
 {\bf 0.247}&
 {\bf 0.262}&
 {\rm 0.289}&
 {\tt 0.336}&
 &           
 &           
 &           
 &           
 &           
 &           
 &           
 &           
 {\tt      }&
 {\tt      }&
           
 \\
 21.00&
 {\bf 0.247}&
 {\bf 0.268}&
 {\it 0.297}&
 {\tt 0.287}&
 &           
 &           
 &           
 &           
 &           
 &           
 &           
 {\tt      }&
 {\tt      }&
 {\tt      }&
 {\tt      }
 \\
 21.10&
 {\bf 0.252}&
 {\bf 0.275}&
 {\rm 0.295}&
 {\rm 0.316}&
 &           
 &           
 &           
 &           
 &           
 &           
 &           
 &           
 {\tt      }&
 {\tt      }&
           
 \\
 21.20&
 {\bf 0.258}&
 {\rm 0.280}&
 {\rm 0.307}&
 {\it 0.312}&
 {\tt 0.294}&
 &           
 &           
 &           
 &           
 &           
 &           
 &           
 {\tt      }&
 {\tt      }&
 {\tt      }
 \\
 21.30&
 {\rm 0.262}&
 {\rm 0.282}&
 {\rm 0.314}&
 {\rm 0.329}&
 {\tt 0.331}&
 &           
 &           
 &           
 &           
 &           
 &           
 {\tt      }&
 {\tt      }&
 {\tt      }&
 {\tt      }
 \\
 21.40&
 {\rm 0.266}&
 {\rm 0.292}&
 {\rm 0.318}&
 {\rm 0.335}&
 {\rm 0.366}&
 &           
 &           
 &           
 &           
 &           
 &           
 {\tt      }
 &           
 &           
 &           
 \\
 21.50&
 {\rm 0.271}&
 {\rm 0.296}&
 {\rm 0.322}&
 {\rm 0.345}&
 {\it 0.353}&
 &           
 &           
 &           
 &           
 &           
 &           
 {\tt      }&
 {\tt      }&
 &           
 {\tt      }
 \\
 21.60&
 {\rm 0.280}&
 {\rm 0.303}&
 {\rm 0.327}&
 {\rm 0.343}&
 {\rm 0.366}&
 {\it 0.391}&
 &           
 &           
 &           
 &           
 &           
 {\tt      }&
 &           
 {\tt      }&
           
 \\
 21.70&
 {\rm 0.295}&
 {\rm 0.313}&
 {\rm 0.335}&
 {\rm 0.353}&
 {\rm 0.370}&
 {\it 0.400}&
 &           
 &           
 &           
 &           
 &           
 &           
 &           
 {\tt      }&
 {\tt      }
 \\
 21.80&
 {\rm 0.303}&
 {\rm 0.319}&
 {\rm 0.337}&
 {\rm 0.353}&
 {\rm 0.375}&
 {\it 0.409}&
 {\it 0.422}&
 {\it 0.421}&
 &           
 &           
 &           
 &           
 {\tt      }&
 {\tt      }&
            
 \\
 21.90&
 {\rm 0.305}&
 {\rm 0.320}&
 {\rm 0.342}&
 {\rm 0.361}&
 {\rm 0.382}&
 {\rm 0.414}&
 {\rm 0.432}&
 {\it 0.458}&
 &           
 &           
 &           
 {\tt      }&
 {\tt      }&
 &           
 {\tt      }
 \\
 22.00&
 {\rm 0.318}&
 {\rm 0.330}&
 {\rm 0.343}&
 {\rm 0.364}&
 {\rm 0.388}&
 {\rm 0.409}&
 {\rm 0.432}&
 {\rm 0.450}&
 {\tt 0.469}&
 {\tt 0.476}&
 &           
 &           
 &           
 &           
 {\tt      }
 \\
 22.10&
 {\rm 0.319}&
 {\rm 0.334}&
 {\rm 0.351}&
 {\rm 0.371}&
 {\rm 0.396}&
 {\rm 0.419}&
 {\rm 0.440}&
 {\rm 0.455}&
 &           
 {\tt 0.466}&
 &           
 &           
 {\tt      }&
 &           
 {\tt      }
 \\
 22.20&
 {\rm 0.327}&
 {\rm 0.341}&
 {\rm 0.357}&
 {\rm 0.377}&
 {\rm 0.398}&
 {\bf 0.421}&
 {\rm 0.444}&
 {\rm 0.458}&
 {\bf 0.466}&
 {\tt 0.471}&
 {\tt 0.479}&
 &           
 {\tt      }
 &           
 &           
 \\
 22.30&
 {\rm 0.337}&
 {\rm 0.346}&
 {\rm 0.363}&
 {\rm 0.382}&
 {\rm 0.404}&
 {\rm 0.425}&
 {\bf 0.439}&
 {\rm 0.452}&
 {\bf 0.471}&
 {\rm 0.475}&
 {\tt 0.488}&
 &           
 &           
 &           
 {\tt      }
 \\
 22.40&
 {\rm 0.343}&
 {\rm 0.354}&
 {\rm 0.365}&
 {\rm 0.388}&
 {\rm 0.403}&
 {\bf 0.427}&
 {\bf 0.444}&
 {\rm 0.463}&
 {\bf 0.471}&
 {\it 0.480}&
 {\tt 0.478}&
 {\tt      }&
 &           
 {\tt      }&
 {\tt      }
 \\
 22.50&
 {\rm 0.352}&
 {\rm 0.361}&
 {\rm 0.374}&
 {\rm 0.391}&
 {\rm 0.410}&
 {\rm 0.429}&
 {\bf 0.448}&
 {\bf 0.458}&
 {\rm 0.473}&
 {\bf 0.485}&
 {\it 0.490}&
 {\tt 0.505}&
 &           
 &           
            
 \\
 22.60&
 {\rm 0.360}&
 {\rm 0.367}&
 {\rm 0.382}&
 {\rm 0.395}&
 {\rm 0.416}&
 {\bf 0.435}&
 {\bf 0.450}&
 {\bf 0.463}&
 {\rm 0.476}&
 {\it 0.489}&
 {\it 0.496}&
 &           
 &           
 &           
            
 \\
 22.70&
 {\rm 0.364}&
 {\it 0.363}&
 {\rm 0.381}&
 {\rm 0.402}&
 {\rm 0.420}&
 {\bf 0.441}&
 {\bf 0.456}&
 {\bf 0.466}&
 {\bf 0.472}&
 {\bf 0.489}&
 {\bf 0.500}&
 {\it 0.509}&
 {\tt 0.509}&
 &           
            
 \\
 22.80&
 {\it 0.380}&
 {\rm 0.373}&
 {\rm 0.393}&
 {\rm 0.410}&
 {\rm 0.421}&
 {\bf 0.445}&
 {\rm 0.460}&
 {\bf 0.468}&
 {\bf 0.481}&
 {\it 0.485}&
 {\rm 0.503}&
 &           
            
 {\tt 0.523}&
 {\tt      }&
 \\
 22.90&
 {\it 0.358}&
 {\rm 0.386}&
 {\rm 0.403}&
 {\rm 0.415}&
 {\rm 0.429}&
 {\rm 0.451}&
 {\bf 0.455}&
 {\bf 0.472}&
 {\bf 0.483}&
 {\bf 0.494}&
 {\rm 0.504}&
 {\bf 0.514}&
 {\tt 0.524}&
 {\tt 0.535}&
            
 \\
 23.00&
 {\rm 0.397}&
 {\it 0.388}&
 {\it 0.405}&
 {\rm 0.419}&
 {\rm 0.439}&
 {\rm 0.451}&
 {\bf 0.464}&
 {\bf 0.478}&
 {\bf 0.484}&
 &           
 {\bf 0.505}&
 {\bf 0.523}&
 {\tt 0.548}&
 &           
            
 \\
 23.10&
 {\it 0.401}&
 {\rm 0.398}&
 {\rm 0.413}&
 {\rm 0.429}&
 {\rm 0.445}&
 {\rm 0.458}&
 {\bf 0.468}&
 {\bf 0.483}&
 {\it 0.484}&
 {\bf 0.502}&
 {\it 0.512}&
 {\tt 0.528}&
 &           
 {\tt 0.515}&
            
 \\
 23.20&
 &           
 {\rm 0.405}&
 {\rm 0.424}&
 {\it 0.429}&
 {\rm 0.449}&
 {\rm 0.463}&
 {\it 0.474}&
 {\it 0.488}&
 {\bf 0.489}&
 {\bf 0.509}&
 {\tt 0.500}&
 &           
 {\tt 0.537}&
 {\tt 0.539}&
            
 \\
\enddata
\tablecomments{The fonts indicate the measured errors in each cell,
where $\sigma_1$ is the estimate of $\sigma_{{\Delta z}/{(1+z)}}$
obtained at the 68\%\ point in the error distribution and $\sigma_2$
is obtained at the 95\%\ point.  {\bf Bold:}
$\sigma_1<0.025,\sigma_2<0.035$; {\rm Roman:} $\sigma_1<0.035,
\sigma_2<0.05$; {\it Italic:} $\sigma_1<0.05, \sigma_2<0.07$; {\tt
Typewriter:} $\sigma_1<0.035,\sigma_2$ undefined (fewer than 30
objects in the bin).}
\end{deluxetable}

\end{document}